\documentclass[onecolumn,showpacs,preprintnumbers,amsmath,preprint]{revtex4-1}

\usepackage{xspace}
\usepackage{epsfig}
\usepackage{amsfonts}
\usepackage{amsmath}
\usepackage{graphicx}
\usepackage{url}
\usepackage[usenames, dvipsnames]{color}
\usepackage{afterpage}

\newcommand{\blue}[1]{\textcolor{black}{#1}}

\title{Final figures for the stanene paper}

\begin{document}


\author{E. G. Marin}
\thanks{These authors contributed equally to this work.}
\author{D. Marian} 
\thanks{These authors contributed equally to this work.}
\author{G. Iannaccone}
\author{G. Fiori}
\email{email: gfiori@mercurio.iet.unipi.it}
\affiliation{Dipartimento di Ingegneria dell'Informazione, Universit\`{a} di Pisa, Pisa, 56122, Italy.}

\title{First principle investigation of Tunnel FET based on nanoribbons from topological two-dimensional material}

\begin{abstract} 
We explore nanoribbons from topological two-dimensional stanene as channel material in tunnel 
field effect transistors. 
This novel technological option offers the possibility to build pure one-dimensional (1D) channel devices \blue{{(comprised of a 1D chain of atoms)}} due to 
localized states in correspondence of the nanoribbon edges.
The investigation is based on first-principle calculations and multi-scale transport simulations
to assess devices performance against industry requirements and their robustness 
with respect to technological issues like line edge roughness, detrimental for nanoribbons. 
{We will show that edges states are robust with respect to 
	the presence of non-idealities (e.g., atoms vacancies at the edges), and that 1D-channel TFETs 
	exhibit interesting potential for digital applications and room for optimization in 
	order to improve the $I_{\rm ON}/I_{\rm OFF}$ at the levels required by the ITRS,}
while opening a path for the exploration of new device concepts at the 
ultimate scaling limits.

\end{abstract}

\maketitle

\newpage

\section*{Introduction}

The isolation of two-dimensional (2D) flakes of graphene in 2004~\cite{Novoselov04} started a revolution,
that has led to the exploration of many other 2D-materials, including Transition Metal Dichalcogenides, 
group III-chalcogenides, groups IVa and Va layered semiconductors~\cite{MXu13}, with physical properties essentially 
different with respect to their three-dimensional (3D) counterparts~\cite{Novoselov16}. The possibility to exploit these properties 
in many-body physics~\cite{Li16}, or in electronics and photonics applications~\cite{Fiori14,Srivastava15} are just some 
of the appeals that have fuelled the research in this area. 

The family of group IVa layered semiconductors and in particular silicene~\cite{Lay12}, germanene~\cite{Davila14} and lastly stanene \cite{Zhu16} deserves 
special mention~\cite{Molle17}. Differently to graphene, these materials show a slightly buckled structure, 
due to their larger interatomic distances, which in turns leads to strong spin-orbit interactions (SOI) 
and unveils the possibility of having topological phases, localized states, and 
ferromagnetic/antiferromagnetic nature~\cite{Le14,Mathes14,Xiong16}, especially
when cut in nanoribbons.

In particular, armchair stanene nanoribbons exhibit a linear dispersion relation~\cite{Houssa16}, 
like silicene or germanene~\cite{Cahangirov10}, but states are not localized at the edges~\cite{Mathes14}  
(see Fig.~S1a in Supplementary Note 1).

On the other hand, zig-zag nanoribbons do show localized states edges with a non-linear energy dispersion,
and have a magnetic nature~\cite{Jung09,Le14,Mathes14,Wang16}, which eventually  
leads to the formation of a bandgap.  
In particular, magnetism in the nanoribbons is predicted to be very dependent 
on the spin correlation length~\cite{Yazyev10,Jung09}, which, at room temperature ($T=300$K) 
is expected to be $\sim$1~nm~\cite{Yazyev10,Jung09}. This means that nanoribbons with width 
smaller than the correlation length exhibit strong exchange interaction between the edges, preserving the magnetism. 

Differently to non-magnetic configurations (where semi-metallic bandstructures are observed~\cite{Wang16}), 
in antiferromagnetic stanene nanoribbons the strong SOI can lead to the opening of a bandgap 
of several hundreds of meV~\cite{Xiong16} (differently to germanene~\cite{Mathes14} and silicene~\cite{Le14}, where the 
bandgap is at most of few tens of meVs), that could open the path towards the exploitation in electronic applications.

A previous attempt to exploit stanene nanoribbons with semi-metallic states in a transistor leveraged the possibility of switching off the current  through 
the scattering modulation when the current is due to non-protected states~\cite{Vandenberghe17}. 
This ingenious proposal could however face some difficulties due to the small range of energies 
where the topological edge states are actually present (see Fig.~S1b in Supplementary Note 1), while possibly being affected by disturbs.

Here, we theoretically explore 1D edge channels \blue{{(comprised of a 1D chain of atoms)}} in Tunnel Field Effect Transistors (TFETs) based on a stanene zig-zag
nanoribbon with an antiferromagnetic ordering. \blue{{Several architectures of TFETs based on 2D materials have been investigated so far \cite{Zhao13,Britnell12,Banerjee09,Sarkar15,Li14,Das14,Roy15}, with quite distinct operational principles (ranging from a theoretical proposal based on electron-hole condensation \cite{Banerjee09} to experimental van-der-Waals heterojunctions \cite{Sarkar15,Roy15}) showing heterogeneous performance. The present manuscript contributes to the efforts in this line with the first demonstration of a TFET based on a topological insulator, with localized channels at the edges.}}
To this purpose,  we employ a multiscale approach~\cite{Fiori13,VIDES}, combining different levels of physical description. 
In particular, Density Functional Theory (DFT) is used to calculate the electronic properties of the stanene nanoribbon, which
are then exploited to define a tight-binding-like Hamiltonian expressed on a Maximally Localized Wannier Functions (MLWF) 
basis set~\cite{Wannier,Pizzi16}. 
The MLWF approach fully retains the first-principles calculations accuracy, 
keeping the localization of the edge states, and reducing the computational burden. 
The extracted Hamiltonian is then used to study  1D edge channels TFET 
under the non-equilibrium Green functions formalism \cite{Datta00} (solved self-consistently with 3D Poisson equation), 
and to explore the robustness of 1D channels against defects and the TFET performance against industry requirements.

\section*{Stanene Zig-zag antiferromagnetic nanoribbon}

\begin{figure} [htp]
\includegraphics[width=0.4\textwidth]{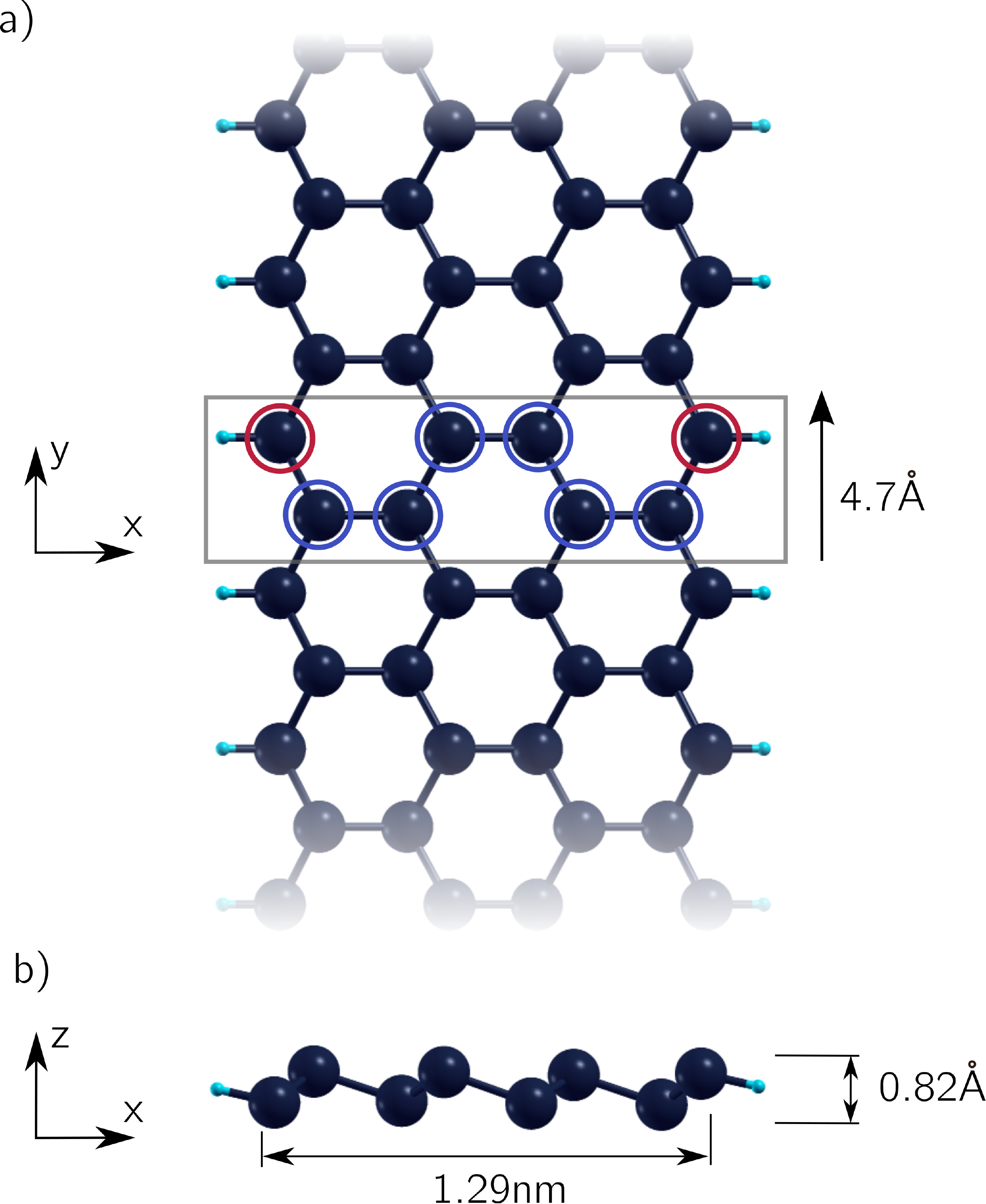}
\caption{\textbf{Geometric structure of the stanene nanoribbon.} A zig-zag configuration of width $1.29$ nm, passivated at the edges with H atoms, is considered. a) Top view. The elementary cell is boxed by a black rectangle. The edge atoms are marked with red circles, and the central atoms are marked with blue circles. b) Lateral view of the nanoribbon. The buckling distance between the two planes of stanene is $0.82$ \r{A}.}
\label{fig:Fig0}
\end{figure}

We have studied the electronic band structure of stanene antiferromagnetic zig-zag nanoribbons (ZZ-NR) 
through Density Functional Theory calculations by means of the Quantum Espresso suite v. 5.2.0~\cite{QE}. 
The geometry of the nanoribbon is shown in Fig.~\ref{fig:Fig0}, with the unit cell embedded in a grey rectangle. 
We have considered a ZZ-NR with a width of $1.29$ nm, 
which is expected to preserve 
edge magnetization consistently with the spin correlation length at room temperature calculated in~\cite{Jung09,Yazyev10}. 
The ZZ-NR is passivated at the edges with H atoms, with a Sn-H bond length of $1.77$ \AA{}. 
As mentioned above, spin orbit interaction (SOI) plays an important role in the electronic bandstructure of group IVa 2D materials, 
which has been here taken into account by means of a non-collinear spin polarized calculation. 
Out-of-plane magnetic moments are oriented in opposite directions at both edges (Fig.~\ref{fig:Fig1}), 
and the in-plane magnetization is negligible. Edge atoms (defined by red circles in Fig.~\ref{fig:Fig0}a) have a strong spin polarization, 
which decreases sharply toward the central atoms (defined by blue circles in Fig.~\ref{fig:Fig0}a).
 The absolute magnetization of the nanoribbon is $0.79\mu_B$/cell (where $\mu_B$ is the Bohr magneton), 
 and the total magnetization is zero, as corresponds to an antiferromagnetic ordering. 

In other IVa 2D materials like graphene, germanene and silicene the origin of the antiferromagnetism in ZZ-NR  has been related to the interactions of localized edge states~\cite{Jung09,Le14,Mathes14}. We have checked the presence of these states in stanene by calculating the normalized contribution of the edge (Fig.~\ref{fig:Fig2}a) and central (Fig.~\ref{fig:Fig2}b) atoms to the density of states (DOS) projected on the bands.
The conduction band minimum (CBM) is located at the $X$ point of the 1D Brillouin zone, and the valence band 
maximum (VBM) is located at $k\approx 0.4\pi/a$ between the $\Gamma$ and $X$ points, resulting in a indirect gap ($\sim0.25$ eV).
States close to the CBM and VBM are completely localized at the edges of the nanoribbon (Fig.~\ref{fig:Fig2}), while the central atoms contribute to the bands elsewhere (Fig.~\ref{fig:Fig2}). 
As can be seen, both the conduction band (CB) and the valence band (VB) have \blue{{high effective masses of $1.8 m_0$ and $1.62 m_0$ for electrons and holes, respectively,}} leading to a large DOS at the edges  (see Fig.~S4 in Supplementary Note 4), which can be interpreted as 1D channels for carrier transport.

To check this idea, and to reduce the computational requirements, we have expressed the Hamiltonian of the system on 
a Maximally Localized Wannier Functions (MLWF) basis set as in~\cite{VIDES}, exploiting
the Wannier90 code~\cite{Wannier}. In particular, we have focused on the first four bands (per spin component) above and below the Fermi level. 
The main contribution to the bonding and anti-bonding states around the gap is due to $p$-orbitals, therefore the 
basis transformation from Bloch states obtained from DFT to MLWF has been accomplished by initially projecting the states on the $p_z$-orbitals 
of the $8$ Sn atoms of the unit cell, resulting into $8$ Wannier functions (per spin component) centered in their respective Sn atoms. As can be 
seen, the tight-binding-like Hamiltonian in the MLWF basis provides an accurate description of stanene ZZ-NR energy bands (see Fig. 
S2 in Supplementary Note 2).

\section*{TFET with 1D-edge channels}

In the light of the previous discussion, we aim at investigating the application of the stanene ZZ-NR in 1D TFET electronic
device~\cite{Jena13,Seabaugh10}.

To this purpose, full 3D simulations have been performed, considering the device shown in Fig.~\ref{fig:Fig1b}. 
In particular, the device is a double gate Tunnel Field Effect Transistor (TFET), with top, bottom, and lateral oxides, 
with thicknesses \blue{{given by}} $t_{\text{ox}}$ and $t_{\text{lat}}$, respectively. If not differently stated $t_{\text{ox}}=0.5$ nm and $t_{\text{lat}}=1$ nm \blue{{are considered}}. 
SiO$_2$ has been assumed as the insulator in all cases. Source and drain regions ($11.3$~nm long) are $p$-type and $n$-type doped, 
respectively, with a molar fraction, $N_\text{\text{D/A}}$, equal to $7.2 \times 10^{-3}$. The channel is $16.9$~nm long and the width is equal to $1.29$~nm. 

Transport is assumed to be ballistic providing an upper limit of on the potential performance for the ultimate scaling limits.
 Non-equilibrium Green functions (NEGF)~\cite{Datta00} and the 3D Poisson equation are 
 self-consistently solved to compute charge density, potential distribution, and current 
 through the NanoTCAD ViDES suite~\cite{VIDESwww}. 

Fig.~\ref{fig:Fig5} shows the free charge (\blue{{a}}) and the intrinsic Fermi level (\blue{{b}}) for three different values of the gate voltage, 
$V_\text{G}$, corresponding to: the OFF state ($V_{\text{OFF}}$), defined as $V_{\text{G}}$ for which the minimum current 
is achieved (center panel); and ON state biases, $V^{\pm}_{\text{ON}}$, corresponding to $V_{\text{G}}=V_{\rm OFF}\pm 0.2$~V (bottom and top panel),
respectively. 
The source-drain voltage, $V_{\text{DS}}$, is set to $0.2$V, as required for ultra-low power operations \cite{ITRS}. 
The width of the nanoribbon is highlighted in Fig.~\ref{fig:Fig5} by the $\pm W/2$ ticks.

From the intrinsic Fermi level profiles (Fig. \ref{fig:Fig5}b) we observe that for $V_{\text{G}}=V^{-}_{\text{ON}}$ (top panel) the channel energy nearly aligns with that of the $p$-type source. As a result an abrupt and thin channel-to-drain barrier is formed, allowing holes in the 1D edges to tunnel from the drain CB to the source/channel VB. For $V_{\text{G}}=V^{+}_{\text{ON}}$ (bottom panel), the energy alignment
 is actually the mirror symmetric, i.e., electrons in the 1D edges from the source VB tunnel through the source-to-channel barrier to the 
 channel/drain CB. \blue{{A schematic depiction of the TFET band diagram is also depicted aside the 2D colormap.}}
 
 The OFF state (central panel) corresponds to the case where the channel energy gap is aligned 
 between the source VB and drain CB forming an opaque barrier, drastically reducing the inter-band tunneling. 
 This discussion is supported by the spatial distribution of the transmission coefficient $T$ (Fig. \ref{fig:Fig5}\blue{{c}}). 
 Indeed, for $V_{\text{G}}=V^{\pm}_{\text{ON}}$, the transmission is large (achieving values $\sim 0.5$) only either at the
 channel-to-drain, or at the source-to-channel junctions, while for $V_{\text{G}}=V_{\text{OFF}}$, 
 $T$ is extremely small ($T<1\times10^{-3}$).
 
 Free charge distribution is shown in Fig.~\ref{fig:Fig5}\blue{{a}}, normalized with respect to the 
electron elementary charge, $e^-$ (i.e., positive/negative magnitudes represent electrons/holes). 
As can be clearly seen, carriers are mostly concentrated at the \blue{{1D}} edges, consistently with the previous discussion. 
In fact, both edges contribute to the valence and conduction bands (as indicated in Fig.~\ref{fig:Fig2}), and these two bands show spin up and spin down character (as explained in Supplementary Note 3). 

In Fig.~\ref{fig:Fig5}d, we show the transfer characteristic, $I_{\text{DS}}$ \emph{vs.} $V_{\text{G}}$, both in linear and 
semilogarithmic scale, \blue{{for a $V_{\text{G}}$ step of $2.5$~mV}}.
The metal work function has been tuned in order to obtain $V_{\text{OFF}}=0$~V. 
As can be seen, subthreshold swings (SS), defined as the inverse slope of the $I_{\text{DS}}$ \emph{vs.} $V_{\text{G}}$ curve in semilogarithmic scale, equal to $SS=15$ mV/dec and $SS=18$ mV/dec are obtained for the $n$ and $p$ branches, respectively.
The ON-OFF current ratio, $I_{\text{ON}}/I_{\text{OFF}}$ (where $I_{\text{OFF}}$ and $I_{\text{ON}}$ are 
defined as the currents for $V_{\text{G}}=V_{\text{OFF}}$ and $V_{\text{G}}=V_{\text{ON}}^{\pm}$, respectively), 
is $\sim 10^3$ for the $n$-type operation and $\sim 0.5\times10^3$ for the $p$-type operation, due to the asymmetry of the conduction and
valence bands. \blue{{The SS, the ON current and the supply voltage comply with the ITRS demands. The $I_{\rm ON}/I_{\rm OFF}$ ratio is close to the value required by high performance applications ($10^4$)
and far from the value required for low power applications ($ 5 \times 10^6$). There is however 
apparent room for optimization, for example by engineering the gate length and source and drain doping \cite{Zhang14}. 
In Table \ref{tab:Tab0}, we compare the main figures of merit obtained from the present work and from TFETs 
based on alternative technologies \cite{Hu14}.
It must be noted that differently from the rest of devices compared in Table \ref{tab:Tab0}, the stanene TFET shows purely 1D conduction (through 1D chains of atoms at the edges), what makes its operation markedly distinct.}}

\fboxrule=4pt

\begin{table}[h]
\renewcommand{\arraystretch}{0.7}
\centering
{
	\begin{tabular}{l l l l l l l}
			\hline
			Material/Device & $I_{\rm OFF}$  &  $I_{\rm ON}$~ &  $I_{\rm ON}/I_{\rm OFF}$ &$SS$ & $V_{\rm DD}$ & Ref. \\ 
			 & (nA/$\mu$m)  &  ($\mu$A/$\mu$m) &  (A/A) &(mV/dec) & (V) & \\
			\hline \hline
			Si/NW & $1\times 10^{-4}$ & $>0.74$ &$>7.4\times 10^{6}$&  67 &  1   & R\cite{Luisier10} \\ \hline
			InAs/NW & $1\times 10^{-4}$ & $3.5$ &$3.5\times 10^{7}$&  42 &  0.3   & R\cite{Conzatti12} \\ \hline		
			Si/DG & $1\times 10^{-3}$ & $1.9\times 10^{-3}$ &$1.9\times 10^{3}$&  68 &  0.3   & R\cite{DeMichielis13} \\ \hline
			InSb/NW & $1.5\times 10^{-3}$ & $>70$ &$>4.6\times 10^{4}$&  35 &  0.6   & R\cite{Sylvia12} \\ \hline
			Ge/NW & $1$ & $11$ &$1.1\times 10^{4}$&  41 &  0.3   & R\cite{Luisier10} \\ \hline
			MoTe$_2$/DG & $1.5$ & $12$ &$8\times 10^{3}$&  7 &  0.1   & R\cite{Ghosh13} \\ \hline
			CNT/GAA & $2$ & $2\times 10^{3}$ &$1\times 10^{3}$&  20 &  0.3   & R\cite{Pillai13} \\ \hline
			Bi$_2$Se$_3$/DG & 5 & 55 &$1.1\times 10^{4}$&  48 &  0.2   & R\cite{Zhang14} \\ \hline
			GNR/SG & $1 \times 10^{2}$ & $175$ &$1.75\times 10^{3}$&  18 &  0.1   & R\cite{Bernstein12} \\ \hline
			1D Stanene/DG ($n$/$p$) & $6 \times 10^{2}$ & 294/573  &$(4.9/9.52)\times 10^{2}$&  15/18 &  0.2  & this work \\ 	 		
			\hline
		\end{tabular}\vspace{0.2cm}
	}
\caption{\blue{{Comparison of the main figures of merit of TFETs made of several 2D channel materials. NW, DG, SG and GAA stand for nanowire, double-gate, single-gate and gate-all-around, respectively.}}}
\label{tab:Tab0}
\end{table}

In Figs.~\ref{fig:Fig7},~\ref{fig:Fig8} and~\ref{fig:Fig9}, we present results regarding the I-V characteristics of the stanene nanoribbon
TFET, spanning among the device parameter space, i.e., considering different doping concentrations, 
gate oxide thicknesses and channel lengths, respectively.

In particular, three different values have been considered for the $p$-type/$n$-type doping of the source and the drain, i.e, $N_\text{\text{D/A}}$, in molar fraction, equal to $6.2 \times 10^{-3}$, $7.2 \times 10^{-3}$ and $8.2 \times 10^{-3}$. \blue{{As aforementioned}}, the doping is an important parameter to engineer 
TFET operation, since it determines the alignment of the source and drain energy bands. The extracted I$_{\text{ON}}$/I$_{\text{OFF}}$ ratio for the three devices 
is equal to  $500$ ($334$), $920$ ($450$), and $388$ ($260$) in the $n$-branch ($p$-branch)  
for $N_\text{\text{D/A}}=6.2 \times 10^{-3}$, $7.2 \times 10^{-3}$, and $8.2 \times 10^{-3}$, respectively. 
The small gap of the stanene nanoribbon, makes it specially sensitive to small variations in the doping 
of the source and drain, being $N_\text{\text{D/A}}=7.2 \times 10^{-3}$ the optimal value for the studied geometry.

The oxide thickness (in combination with the channel thickness, $t_{\text{s}}$) determines the scaling length, 
$\sqrt{t_{\text{s}} t_{\text{ox}} (\epsilon_{\text{s}}/\epsilon_{\text{ox}}) }$ \cite{Ferain11}, a quantitative measure of the drop of the gate 
electrostatic potential: the smaller the scaling length, the better  the control of the gate over the channel. In this regard, 
the monoatomic thickness of 2D materials guarantee the minimum $t_{\text{s}}$ achievable. 
Better electrostatic control can then be achieve either reducing $t_{\text{ox}}$ or increasing the dielectric constant 
of the oxide $\epsilon_{\text{\text{ox}}}$. For the 1D-edge channels TFET, the oxide thickness significantly 
reduces the ON current by approximately $\times 1.5$ every $0.5$nm for both the $n$- and $p$ branches 
(Fig.~\ref{fig:Fig8}). 

In Fig.~\ref{fig:Fig9}, we have compared TFET transfer characteristics for three channel lengths, 
$L_{\text{ch}}$, $5.6$ nm, $11.3$ nm, $16.9$ nm and {\blue{{$22.3$~nm}}} in order to explore the short-channel-effects (SCE) influence on the 1D channels performance. 
Indeed, in ultra-scale devices, source and drain contacts introduce large parasitic capacitances, undermining the effect of the gate.
In addition \blue{{inter-band}} tunneling plays a relevant role, affecting the performance in the OFF state.

As can be seen, in the case of the shortest considered channel length,
a strong degradation of both the ON/OFF ratio ($\sim 90$) and the $SS$ ($\sim 190$mV/dec) is observed. 
For $L_{\text{ch}}=11.3$~nm, $SS$ is not much degraded ($\sim 31$mV/dec), pointing out that the gate recovers 
the control over the channel, but the OFF state current is still degraded $\sim 3.6\times$ due to \blue{{inter-band}} tunneling. 
\blue{{For $L_{\text{ch}}=22.3$ nm, $SS$ and the OFF current are no better improved as compared to shorter channel lengths,
indicating that the intrinsic properties of the material (i.e., the bandgap) are limiting the performance, and pointing out the necessity of other gate engineering solutions or scaling of the supply voltage \cite{Zhang14}. }}

\begin{figure} [htp]
\includegraphics[width=0.4\textwidth]{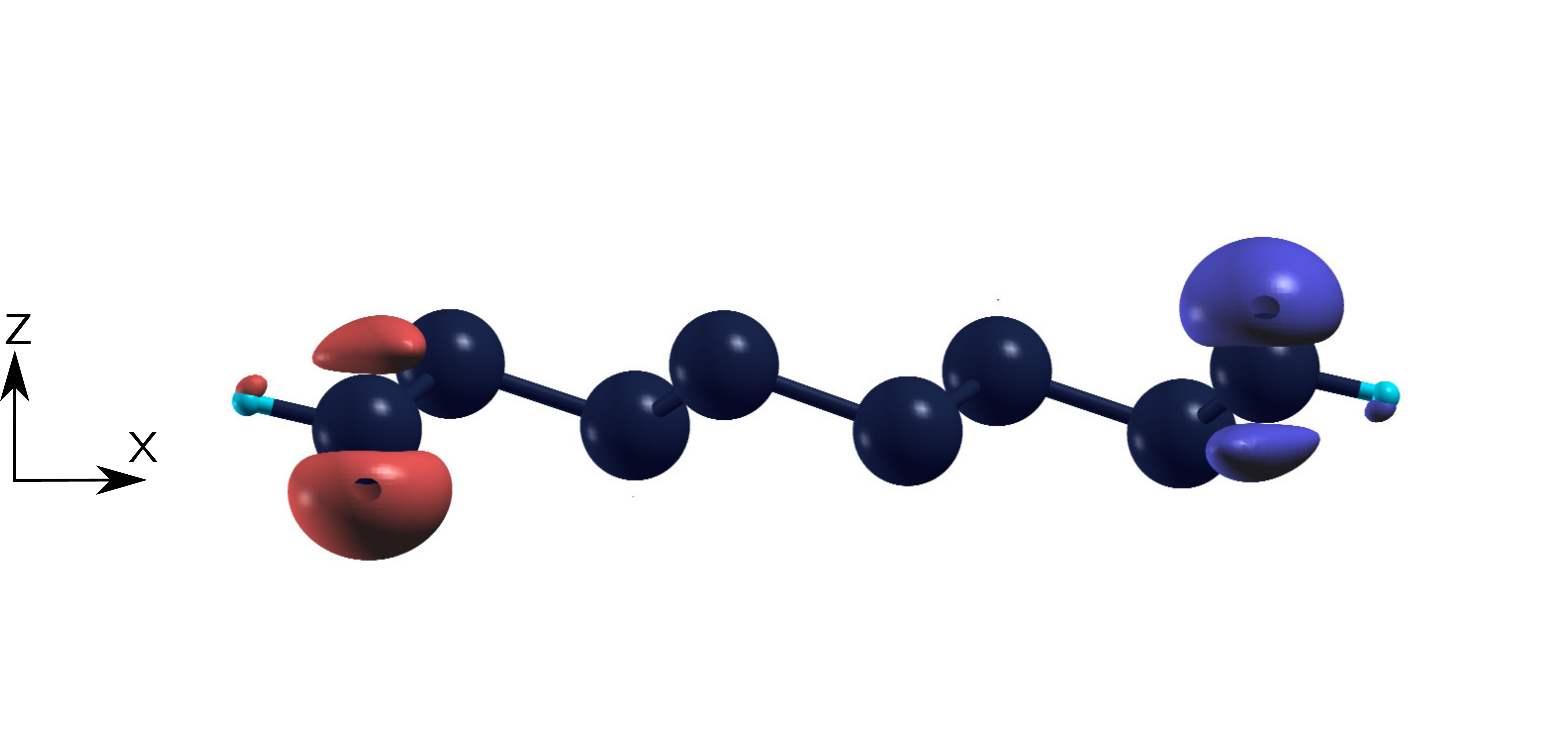}
\caption{\textbf{Magnetization at the edges of the stanene nanoribbon.} Isosurfaces corresponding to a magnetization of $-1.5\times10^{-3}\mu_{\text{b}}$/cell (blue) and $+1.5\times10^{-3}\mu_{\text{b}}$/cell (red) along the $\hat{z}$-direction. Negligible magnetizations along the $\hat{x}-$ and $\hat{y}-$directions are observed. Total magnetization in the cell is zero, and absolute magnetization is $0.79\mu_{\text{B}}/$cell. The spin moments are oriented in opposite direction depending on the edge.}  
\label{fig:Fig1}
\end{figure}

\begin{figure} [tbp]
\includegraphics[width=0.6\textwidth]{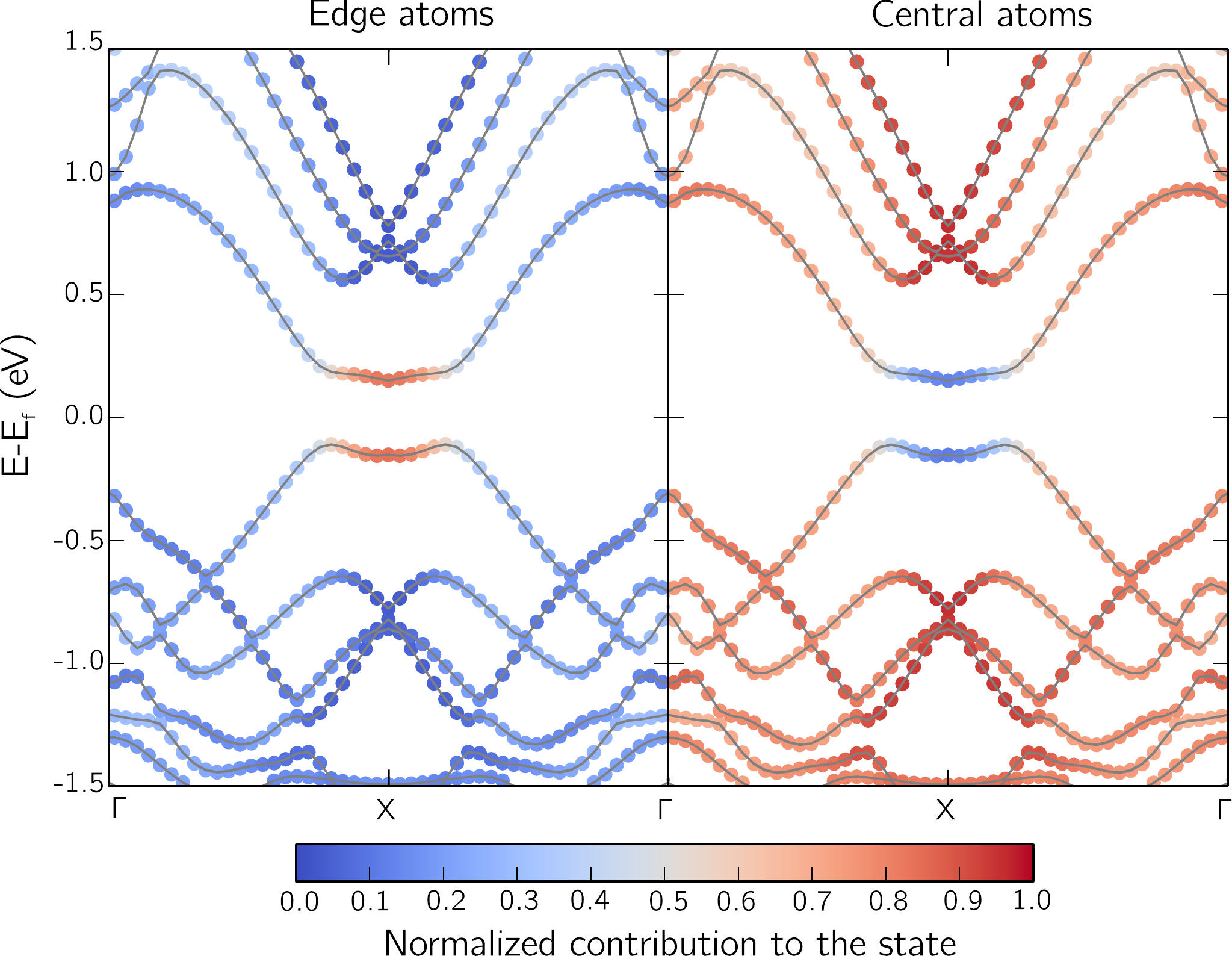}
\caption{\textbf{Density of states projected on the bandstructure.} Normalized contribution of the edge (right) and central (left) atoms to the DOS projected on the bands for the ZZ stanene nanoribbon. The Fermi level is set to 0 eV. The dominant contribution of the edge atoms to the states close to the Fermi level can be clearly observed.
}
\label{fig:Fig2}
\end{figure}

\begin{figure} [tbp]
	\includegraphics[width=0.6\textwidth]{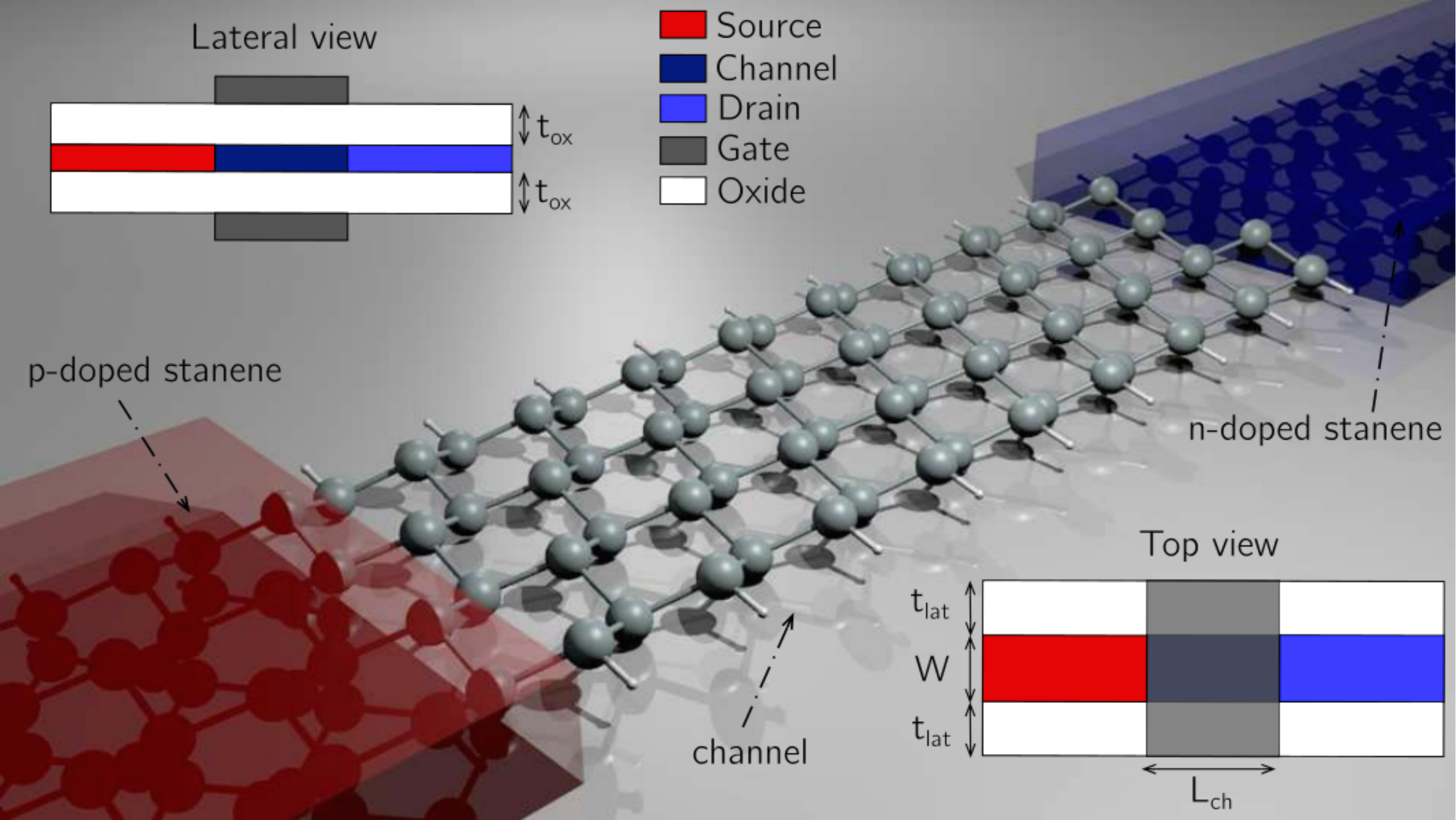}
	\caption{\textbf{Sketch of the 1D-edge-channels TFET.} Upper-left and lower-bottom corner insets show lateral and top views of the geometry. The stanene ZZ-NR is embedded in top, bottom, and lateral oxides, with thicknesses $t_{\text{ox}}$ and $t_{\text{lat}}$. The channel width is $W$ and its length is $L_{\text{ch}}$. A double-gate structure has been considered. The source and drain region are $p$- and $n$-doped, respectively.}
	\label{fig:Fig1b}
\end{figure}

\begin{figure} [tbp]
{\includegraphics[width=0.8\textwidth]{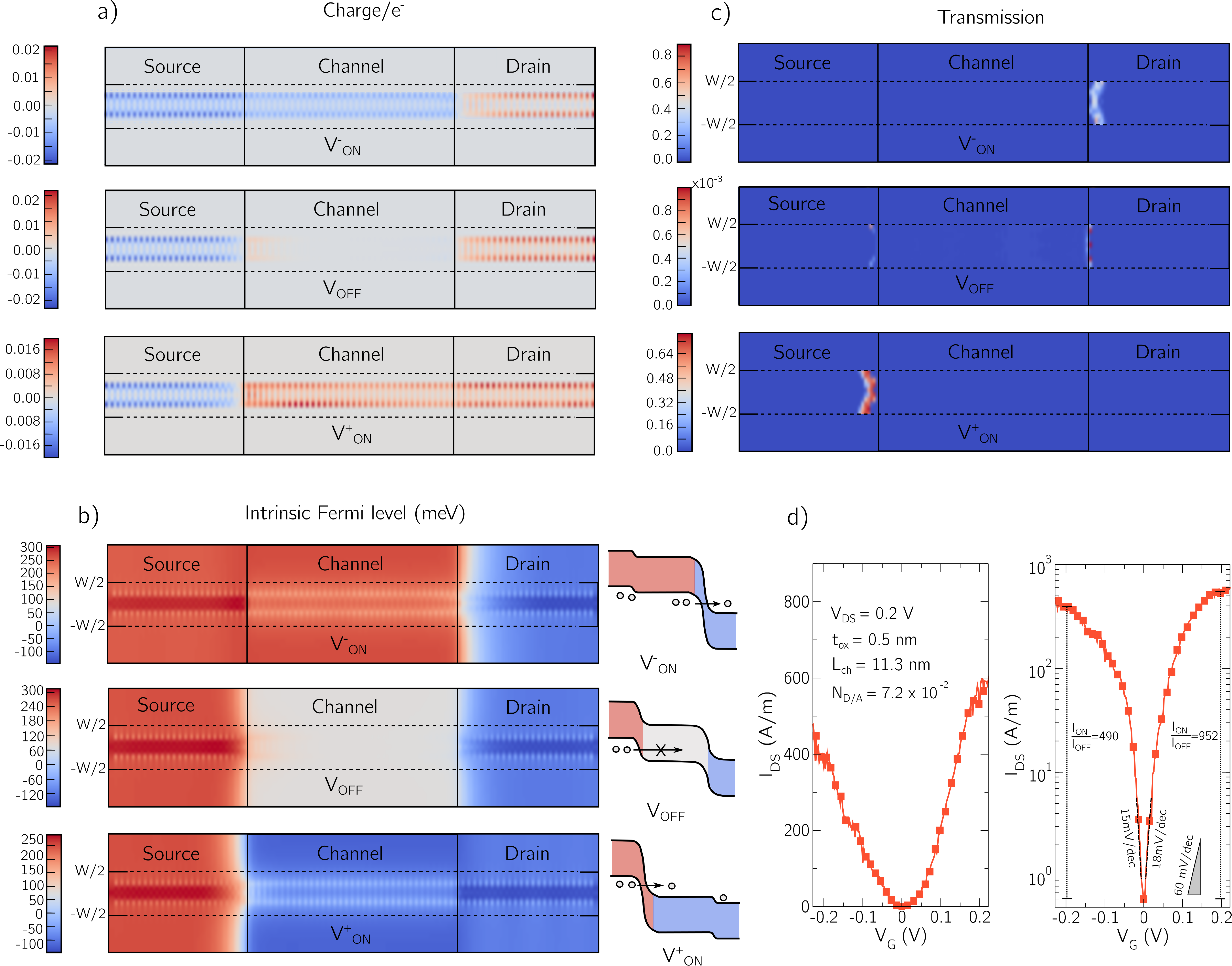}}
\caption{\blue{\textbf{Local intrinsic Fermi level, charge, and transmission coefficient of the stanene nanoribbon.} 2D colormap of a) the free charge distribution and b) the intrinsic Fermi energy, c) transmission coefficient at the intrinsic Fermi energy as a function of the position of the nanoribbon, for three different bias conditions: $p$-conductive ON state (top), OFF state (center) and $n$-conductive ON state. V$_{\text{OFF}} = 0$ V, i.e. the minimum of the transfer characteristic in d). A sketch of the TFET band diagram is depicted aside the intrinsic Fermi level colormap. The actual width of the nanoribbon is marked by dashed lines. A lateral oxide of $1$ nm is considered at both sides of the nanoribbon. d) Transfer characteristic of the TFET in linear and semi-logarithmic scales.}
}
\label{fig:Fig5}
\end{figure}

\begin{figure} [tbp]
\includegraphics[width=0.6\textwidth]{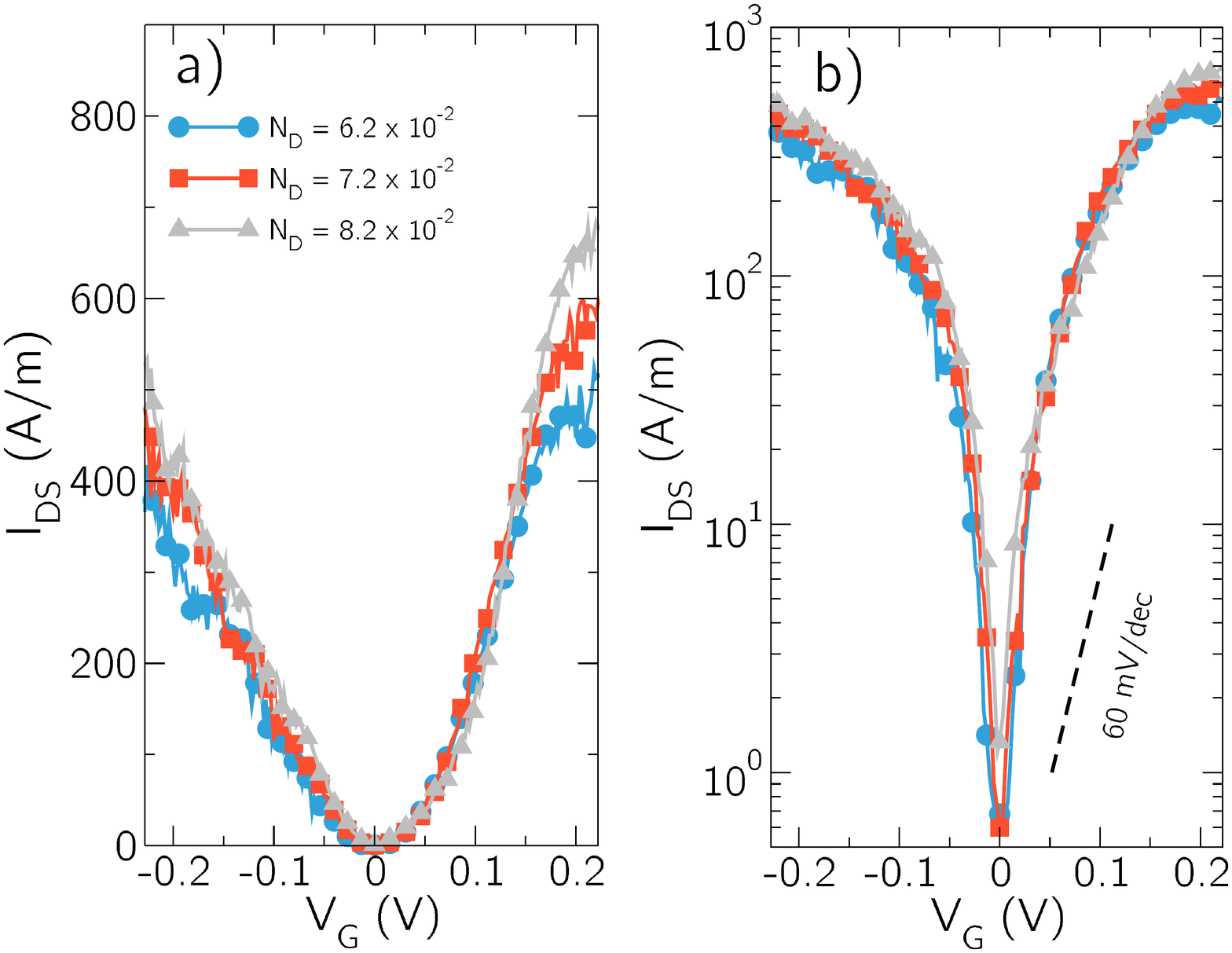}
\caption{\textbf{TFET $I_{\text{DS}}$ \emph{vs.} $V_{\text{G}}$ for different source/drain doping concentrations.} Transfer characteristics in a) linear and b) semi-logarithmic scale of the TFET for different values of doping concentrations at the source/drain: $N_\text{\text{D/A}}$ equal to $6.2 \cdot 10^{-3}$ (blue circles), $7.2 \cdot 10^{-3}$ (red squares), and $8.2 \cdot 10^{-3}$ (gray triangles).}
\label{fig:Fig7}
\end{figure}

\begin{figure} [tbp]
\includegraphics[width=0.6\textwidth]{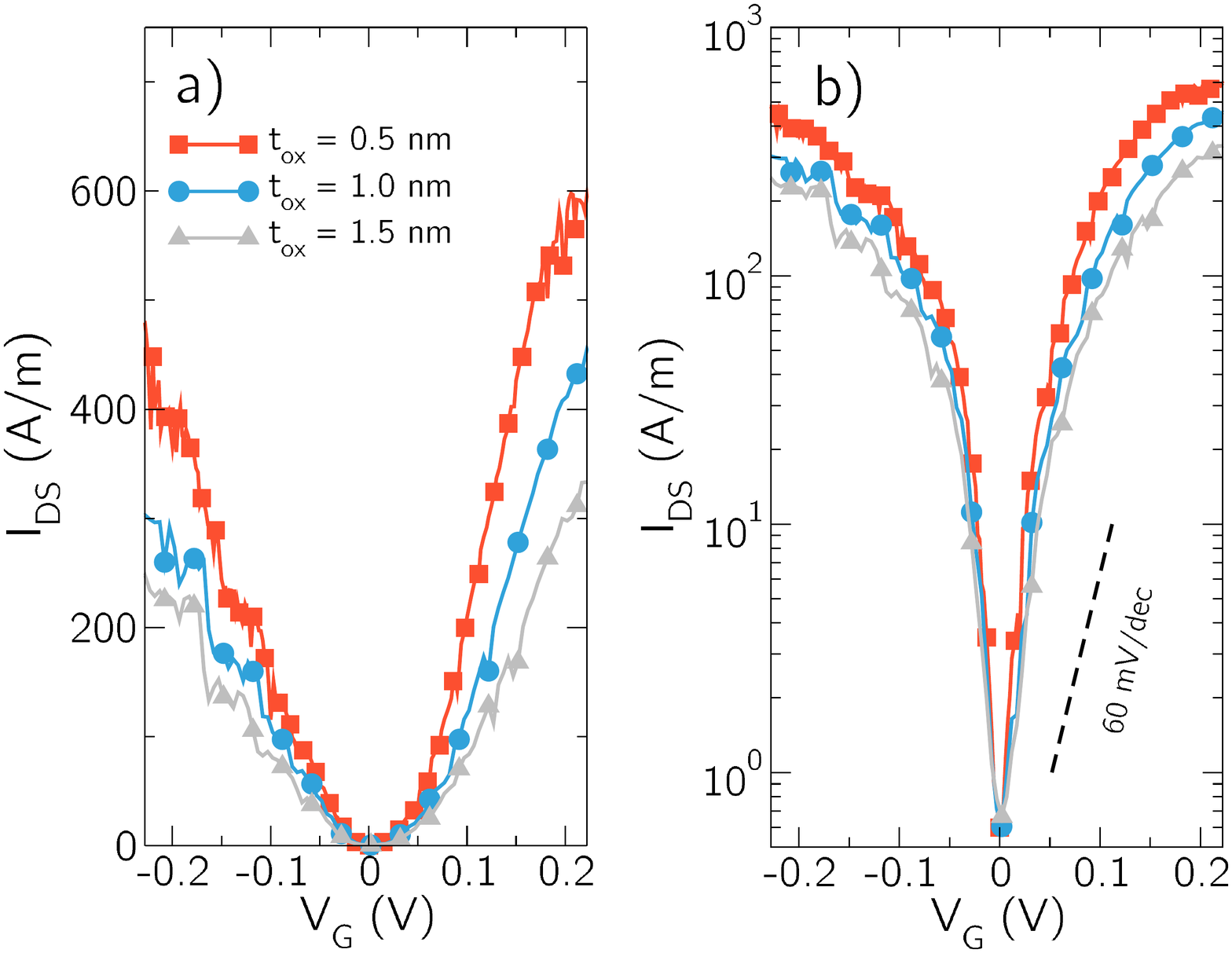}
\caption{\textbf{Impact of oxide scaling on the 1D-channels TFET.} Transfer characteristics in a) linear and b) logarithmic scale of the TFET for different oxide thicknesses, $t_{\text{ox}}$, equal to $0.5$nm (red squares), $1$nm (blue circles), and $1.5$nm (gray triangles).}
\label{fig:Fig8}
\end{figure}

\begin{figure} [tbp]
	{\includegraphics[width=0.6\textwidth]{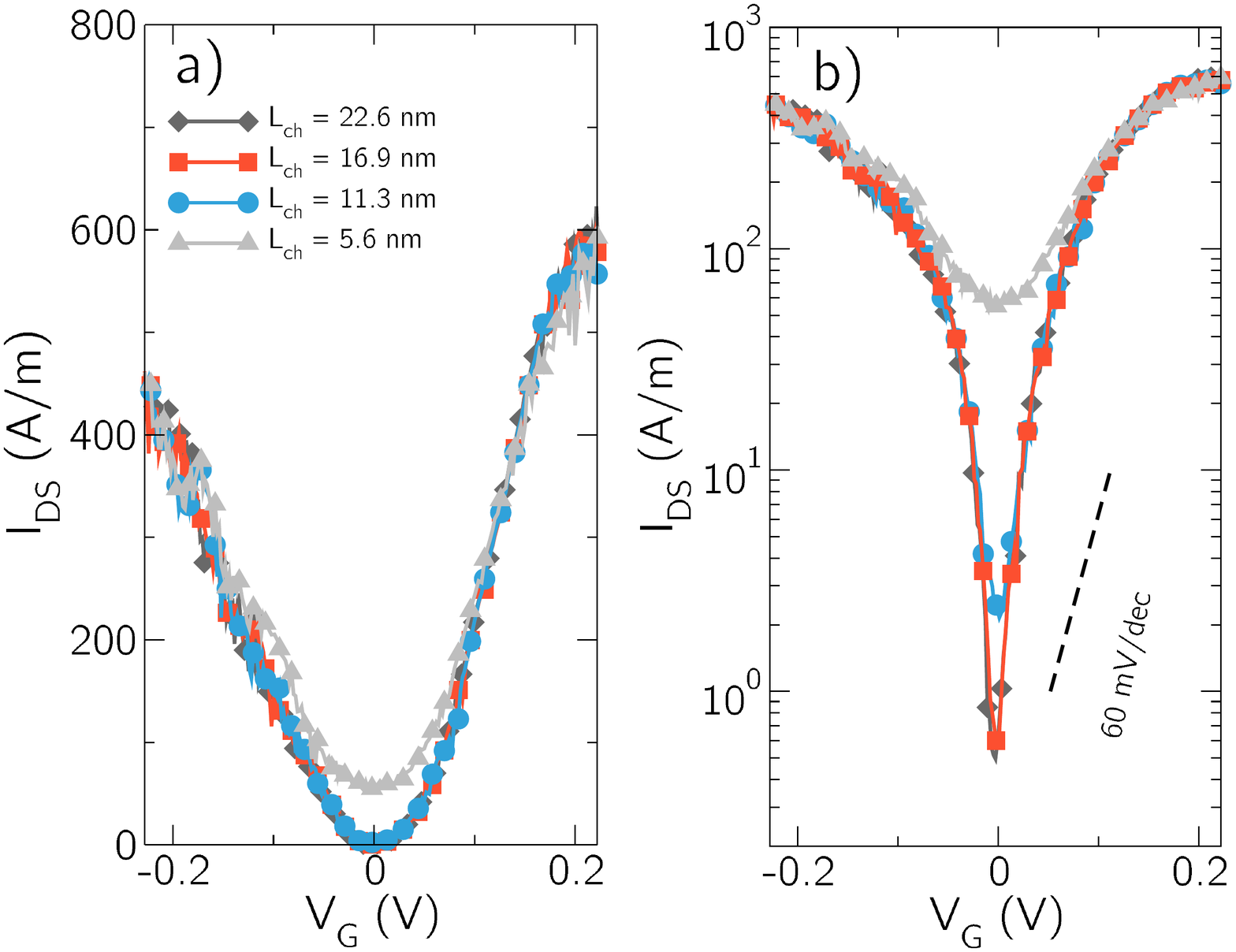}}
	\caption{\blue{\textbf{Impact of channel-length scaling on the 1D-channels TFET.} Transfer characteristics in a) linear and b) logarithmic scale of the TFET for different gate underlaps: $22.6$ nm (black diamonds), $16.9$ nm (red squares), $11.3$ nm (blue circles), and $5.6$ nm (gray triangles).}}
	\label{fig:Fig9}
\end{figure}

We have previously stated that the 1D channels in the stanene ZZ-NR have opposite spin at 
both edges (Fig.~\ref{fig:Fig1} and Fig.~\ref{fig:Fig2}). This might be thought as the main ingredient to expect 
protection against back-scattering. However, this not hold for ZZ-NRs with antiferromagnetic ordering. 
A careful look at the bands and the spins (Supplementary Note 3) shows that transitions from $+k$ to $-k$ 
are possible within the same spin component and therefore the states are not protected~\cite{Manoharan10}.

\begin{figure} [tbp]
	\includegraphics[width=0.6\textwidth]{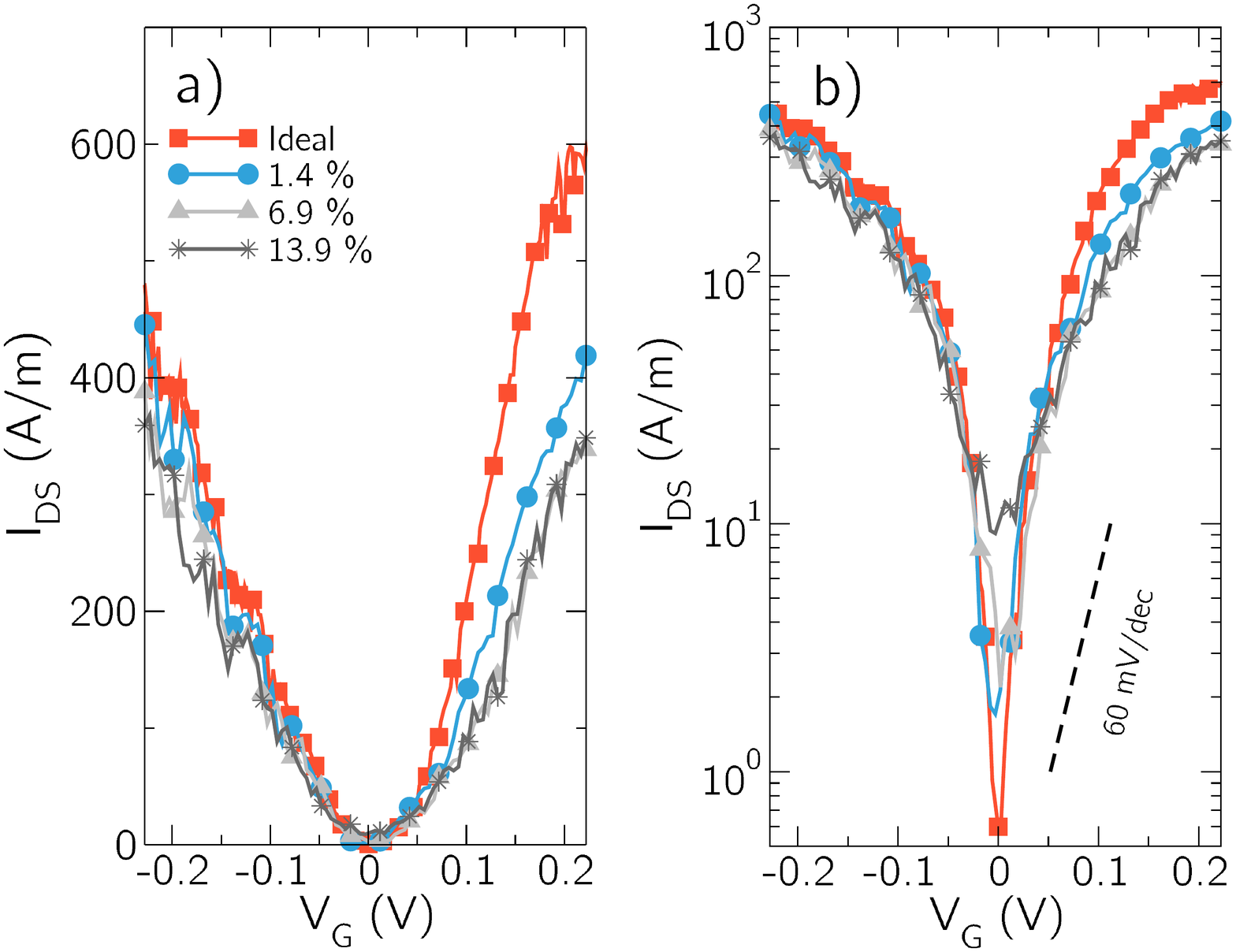}
\caption{\textbf{1D-channels TFET robustness against LER.} Transfer characteristics in a) linear and b) semi-logarithmic scale of the TFET for different percentages of defects at the edges: no defects (red squares), 1.4$\%$ (blue circles), $6.9\%$ (gray triangles) and $13.9\%$ (black asterisks).}
\label{fig:Fig6}
\end{figure}

The 1D channels located at the edges of the NR and the device behavior might be strongly affected 
by line edge roughness (LER) \cite{Fiori07}. 
To check the sensitivity of the transistor operation to LER, we have 
calculated the transfer characteristic for different percentage of defective atoms at the edges (Fig.~\ref{fig:Fig6} and Methods).
As can be seen, despite the large number of defects, I$_{\text{ON}}$ preserved and decreases 
with respect to the ideal case at most by only a factor of $\sim1.8$ ($\sim1.3$) for the $n$-type ($p$-type) branch. 
Transport results are in good agreement with DFT calculations (Fig.~S5 in Supplementary Note 5), 
where we have demonstrated that localized edge states are conserved, 
when different number of edge atoms are removed. 

The less pronounced degradation of the ON current for the $p$-type conductive branch, might be due to the fact that
 VB has a non-negligible contribution of the central atoms around its maxima (Fig.~\ref{fig:Fig1}). 
 The OFF state current, on the contrary, is more affected by the presence of defects increasing 
 by a factor $\sim20$ in comparison with the ideal case, {what as explained in a previous work might be due to appearance of mid-gap states\cite{Yoon08}.}
 The sub-threshold swing $SS$, is consequently affected, shifting from the sub-thermionic values $18$mV/dec ($15$mV/dec) 
 for the ideal case, to the over-thermionic values $99$mV/dec ($75$mv/dec) 
 for the highest percentage of defects in the $n$-branch ($p $-branch). Smaller number of defects 
 still preserves sub-thermionic $SS$ equal to $32$mV/dec ($24$mV/dec) and $44$mV/dec ($38$mV/dec) 
 for $1.4 \%$ and $6.9 \%$ percentages in the $n$-branch ($p$-branch), respectively.

\section*{Conclusion}

The combination of localized edge states and energy gap that arises from the antiferromagnetic ordering in 
group IVa nanoribbons, opens a new path for the exploration of electronic devices. We have investigated 
tunneling field effect transistors based on stanene zig-zag nanoribbons, with purely 1D 
channels for both electrons and holes. The potential performance of the 1D channels TFET has been 
assessed by a comprehensive multiscale study combining the predictive power of Density Functional Theory 
calculations, the computational efficiency of tight-binding Hamiltonians obtained by means of Maximum 
Localized Wannier Functions, and full self-consistent device simulations based on non-equilibrium Green 
functions and 3D Poisson equation. Our results show that the 1D edge channels are robust 
against line-edge roughness and \blue{{that they show interesting performance against industry requirements, although the small bandgap of stanene demands gate and doping optimization to 
achieve compliant $I_{\rm ON}/I_{\rm OFF}$ ratio.}}.
The proposed study theoretically demonstrate for the first time the full operation of 1D edges TFETs 
based on stanene nanoribbons, opening the path towards 1D ultra-scale devices.

\section*{Methods}

\textbf{Density Functional Theory}. First principle calculations have been performed using Density Functional Theory 
as implemented in Quantum Espresso v. 5.2.0~\cite{QE}. Supercells with $32$ \AA{} and $20$ \AA{} of vacuum  in 
the $\hat{y}$-direction (namely the NR width) and $\hat{z}$-direction (namely orthogonal to the NR plane) were used 
to minimize the interaction between periodic repetitions of the cell. A structural optimization of all geometries was 
performed within the Broyden-Fletcher-Goldfarb-Shanno algorithm until forces were less than $5\times 10^{-3}$eV$/
$\AA{}. The convergence for energies was set to $10^{-6}$ eV. A Perdew-Burke-Ernzerhof exchange-correlation 
functional (Generalized Gradient Approximation) was considered~\cite{PBE} within fully-relativistic norm-conserving 
pseudopotentials including non-collinear corrections. The pseudopotentials were obtained from the SSSP Library, for 
both Sn and H~\cite{SSSP}. Energy cutoffs for charge density and wavefunction expansions were set to $360$ Ry 
and $30$ Ry, respectively. SOC was included in the self-consistent calculations of the charge density. Integration in 
the Brillouin zone was accomplished in a $1\times10\times1$ $\Gamma$-centered grid for the ZZ-NR, whilst a 
$1\times6\times1$ $\Gamma$-centered grid was considered for the 9-unit-cell NRs discussed in the Supporting 
Information. 
The band-structure calculations were performed for a $50$-points $k$-grid on a highly symmetric path, $\Gamma$-$X$-$\Gamma$, along the 1D Brillouin zone.
\vspace{0.5cm}

\textbf{Maximally Localized Wannier Functions}. Wannier90 suite v. 2.0.1~\cite{Wannier} was used to compute the 
Maximally Localized Wannier Functions (MLWFs). The same $k$-sampling of the Brillouin zone as in the DFT 
simulations has been used to compute the overlap matrices required to determine the MLWFs. 
16 bands around the fundamental gap have been considered (including spin degeneration).
A threshold of $10^{-10}$\AA{}$^2$ was set for the total spread change in the MLWFs in $20$ consecutive 
iterations. Initial projection on $p_z$-type orbital centered on each of the $8$ Sn atoms were considered for the 
Wannierisation. The band-structure was then calculated on a denser $k$-grid, $1\times100\times1$, along the 
same $\Gamma$-$X$-$\Gamma$ path as in DFT.
\vspace{0.5cm}

\textbf{Transport simulations}. Device transport simulations were performed with the NanoTCAD ViDES code~\cite{VIDES,VIDESwww}. Non-equilibrium Green functions (NEGF) and three-dimensional (3D) Poisson equation 
were self-consistently solved, \blue{{under varying bias conditions, with a $V_{\rm G}$ step of $2.5$~mV}}, determining the device electrostatic potential, charge density, and current in the 
ballistic regime at room temperature ($T=300$ K). The MLWF Hamiltonian with up to 96 nearest-neighbors was 
used to compute NEGF within a energy step of $1$ meV and a energy window ranging from $-1.5$ eV to $1.5$ eV 
(below/above the Fermi level at the source). For Poisson, the mean position of the MLWF was taken as 
representative of the location of the charge center. In order to deal with a reasonably discretized grid, charge centers 
were grouped, inside the unit cell, into $4$ grid points along the $x$-direction with squeezed $y$ and $z$ 
coordinates. A resulting uniform $11\times168\times25$ grid was used in the 3D Poisson for the channel length of 
$16.9$ nm. Convergence in the self-consistent calculation was accomplished when the change in the potential at 
any grid point was less than $10^{-2}$ V. Line edge roughness was simulated by introducing arbitrary defects along 
the edges of the NR channel. The on-site energies of the Hamiltonian corresponding to MLWF where a defect was 
introduced were forced to a large constant value of energy \blue{{emulating this way the presence of a barrier. An additional check consist of setting the coupling terms to this site in the Hamiltonian to zero, thus removing the possibility of hopping. This methodology leads to equivalent results (see Supplementary Information) but it is less stable in terms of convergence.}}

\section*{Acknowledgments}
Authors gratefully acknowledge the support from the Graphene Flagship (No. 696656).

\section*{Supplementary information}
Supplementary Information accompanies this paper.
\newpage

\fboxrule=4pt

\clearpage
\pagebreak

\end{document}